\documentclass{pasj00}
\draft

\begin{document}
\SetRunningHead{Honma et al.}{First Fringe Detection with VERA}
\Received{2002/12/17}
\Accepted{}

\title{First Fringe Detection with VERA's Dual-Beam System and \\ its Phase Referencing Capability}


\author{
Mareki \textsc{Honma}\altaffilmark{1,2,3}
Takahiro \textsc{Fujii}\altaffilmark{1,6}
Tomoya \textsc{Hirota}\altaffilmark{6}
Koji \textsc{Horiai}\altaffilmark{1,4} \\
Kenzaburo \textsc{Iwadate}\altaffilmark{1,4}
Takaaki \textsc{Jike}\altaffilmark{1}
Osamu \textsc{Kameya}\altaffilmark{1,3,4}
Ryuichi \textsc{Kamohara}\altaffilmark{1,6} \\
Yukitoshi \textsc{Kan-ya}\altaffilmark{1}
Noriyuki \textsc{Kawaguchi}\altaffilmark{1,2,3}
Hideyuki \textsc{Kobayashi}\altaffilmark{1,5}
Seisuke \textsc{Kuji}\altaffilmark{1,2} \\
Tomoharu \textsc{Kurayama}\altaffilmark{1,7}
Seiji \textsc{Manabe}\altaffilmark{1,2,3}
Takeshi \textsc{Miyaji}\altaffilmark{1,5}
Kouichirou \textsc{Nakashima}\altaffilmark{6} \\
Toshihiro \textsc{Omodaka}\altaffilmark{6}
Tomoaki \textsc{Oyama}\altaffilmark{1,7}
Satoshi \textsc{Sakai}\altaffilmark{1,2}
Sei-ichiro \textsc{Sakakibara}\altaffilmark{6}\\
Katsuhisa \textsc{Sato}\altaffilmark{1,4}
Tetsuo \textsc{Sasao}\altaffilmark{1,2}
Katsunori M. \textsc{Shibata}\altaffilmark{1,2}
Rie \textsc{Shimizu}\altaffilmark{6}\\
Hiroshi \textsc{Suda}\altaffilmark{1,7}
Yoshiaki \textsc{Tamura}\altaffilmark{1,2,3}
Hideki \textsc{Ujihara}\altaffilmark{1}
Akane \textsc{Yoshimura}\altaffilmark{6}
}

\altaffiltext{1}{VERA Project Office, NAOJ, Mitaka, 181-8588}
\altaffiltext{2}{Earth Rotation Division, NAOJ, Mizusawa, 023-0861}
\altaffiltext{3}{Graduate University for Advanced Studies, Mitaka, 181-8588}
\altaffiltext{4}{Mizusawa Astro-Geodynamics Observatory, NAOJ, Mizusawa, 023-0861}
\altaffiltext{5}{Radio Astronomy Division, NAOJ, Mitaka, 181-8588}
\altaffiltext{6}{Faculty of Science, Kagoshima University, Korimoto, Kagoshima 890-0065}
\altaffiltext{7}{Department of Astronomy, University of Tokyo, Bunkyou, Tokyo 113-8654}

\email{(MH) honmamr@cc.nao.ac.jp}


\maketitle

\begin{abstract}
We present the results of first dual-beam observations with VERA (VLBI Exploration of Radio Astrometry).
The first dual-beam observations of a pair of H$_2$O maser sources W49N and OH43.8-0.1 have been carried out on 2002 May 29 and July 23, and fringes of H$_2$O maser lines at 22 GHz have been successfully detected.
While the residual fringe phases of both sources show rapid variations over 360 degree due to the atmospheric fluctuation, the differential phase between the two sources remains constant for 1 hour with r.m.s. of 8 degree, demonstrating that the atmospheric phase fluctuation is removed effectively by the dual-beam phase referencing.
The analysis based on Allan standard deviation reveals that the differential phase is mostly dominated by white phase noise, and the coherence function calculated from the differential phase shows that after phase referencing the fringe visibility can be integrated for arbitrarily long time.
These results demonstrate VERA's high capability of phase referencing, indicating that VERA is a promising tool for phase referencing VLBI astrometry at 10 $\mu$as-level accuracy.

\end{abstract}

\section{Introduction}

VERA (VLBI Exploration of Radio Astrometry, Sasao 1996; Honma et al. 2000 and references therein) is a new VLBI array dedicated to phase referencing astrometry.
Referring to distant radio galaxies and QSOs, VERA measures the proper motions and parallaxes of Galactic H$_2$O and SiO maser sources (mainly star forming regions and Mira variables) with 10 $\mu$as-level accuracy.
With that high accuracy, VERA can measure parallax of masers located at $D$ kpc away with an accuracy of D\% and proper motion with an accuracy of 0.05$\times D$ km s$^{-1}$ for a monitoring time span of one year.
Hence VERA will be able to establish 3-D structure and dynamics of the Milky Way with unprecedentedly high accuracy (for detailed scientific targets, see Honma et al. 2000).

The key to precise VLBI astrometry at 22 GHz and 43 GHz is how to remove the atmospheric phase fluctuation based on phase referencing.
Previously, phase referencing has been done based on various methods including fast switching, paired/clustered antennas, and water vapor radiometer (e.g., Beasley and Conway 1995; Asaki et al. 1996; 1998; Rioja et al. 1997; Carilli and Holdaway 1999; Thompson et al. 2001).
Among them the most popular one is fast switching, in which two adjacent sources are observed in turn in a short cycle (typically $\sim$ 1 minute).
There have already been some examples of switching VLBI observations at 22GHz and 43 GHz such as QSO-QSO astrometry by Guirado et al.(2000) and proper motion measurement of Sgr A* by Reid et al.(1999), but achieved astrometric accuracy remained around 100 $\mu$as.

In order to achieve higher performance in phase referencing and also in astrometric measurement, VERA utilizes a rather unique system, so-called 'dual-beam' antenna in which two steerable receivers located at the focal plane simultaneously observe two sources (target and reference sources) with separation ranging from 0.3 to 2.2 degrees (Kawaguchi et al. 2000).
With dual-beam observation, the phase fluctuations of two sources can be measured at the same time, and thus one can expect higher performance in phase referencing.

In order to investigate how effectively VERA's dual-beam system works for phase referencing, we have been conducting test observations since the completion of all four stations in early 2002.
Here we report the results of first fringe observations with dual-beam system, and demonstrate its high capability of phase referencing.

\section{Observation and Reduction}

\begin{figure}
\begin{center}
        \FigureFile(15cm,15cm){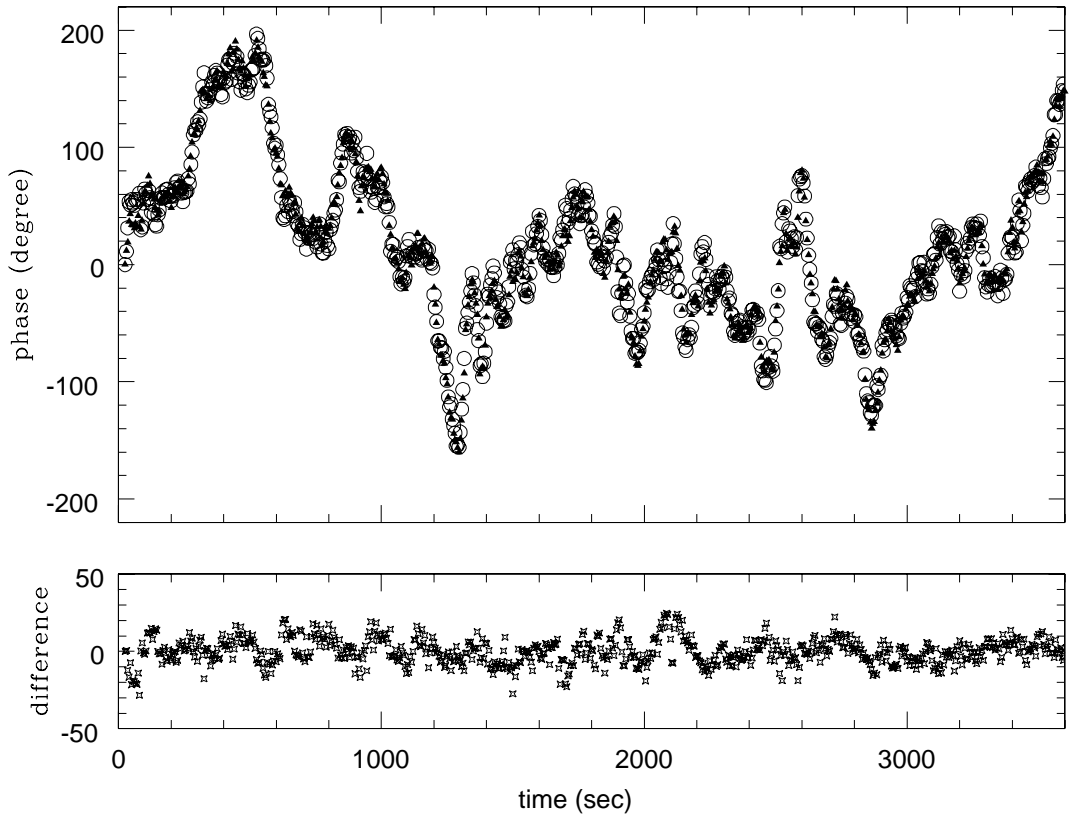}
\end{center}
\caption{(top) : Residual fringe phases for W49N (open circle) and OH43.8-0.1 (triangle) intensity peak channel. (bottom) : differential fringe phase (phase difference of the two sources).
Time origin is UT 11:30 of 2002 July 23.
}
\end{figure}

After the detection of single-beam first fringe of ORI-KL H$_2$O maser with Mizusawa-Iriki baseline on 2002 February 20, we made the first dual-beam VLBI observation with VERA on 2002 May 29 at 22 GHz.
A pair of strong H$_2$O maser sources W49N and OH43.8-0.1, separated by 0.65 degree on the sky plane, was observed with Mizusawa and Iriki stations from UT 13:15 for about 40 minutes.
The sky conditions at two stations were moderate, with system noise temperatures (at the zenith) between 300 and 400 K at both stations.
Only left-hand circular polarization was received and sampled with 2-bit quantization, and filtered with bandwidth of 16 MHz for each source using VERA Digital Filter before being recorded onto magnetic tapes at the rate of 128 Megabit s$^{-1}$ (16MHz band $\times$ 2 beams).
Correlation process was carried out on Mitaka FX correlator located at NAOJ Mitaka campus with spectral resolution of 31.25 kHz (512 channels per 16 MHz band), and fringes were successfully detected for both W49N and OH43.8-0.1.
In order to confirm the system reliability, we also made the dual-beam observation of the same sources on 2000 July 23 for a longer period with the same system configuration.
The dual-beam fringes were successfully detected 
again for the second observation.

Due to the error in tracking-center positions of the maser sources as well as error in station positions, the residual fringe rates of raw correlator outputs were fairly large for both sources and for both observational sessions in May and July.
To correct for the effect of residual fringe rate and fringe phase, we fitted the fringe phase (and also the fringe rate) based on the relation $\phi=U\Delta X + V\Delta Y$ ($\phi$ is the fringe phase, $U$ and $V$ are the projected baseline components in the UV plane, and $\Delta X$ and $\Delta Y$ are position offsets of tracking center), assuming that the residual fringe phase totally comes from position error of tracking center.
This procedure can be regarded as 'high-pass filtering', in which rapid phase variation due to atmospheric fluctuation is conserved while slowly-varying phase error due to position offset is corrected for.
In the following sections we present the residual fringe phase after the correction of position offset.

For both observations, no calibration was made to correct for variation of instrumental delay difference in dual-beam system.
Recent results of dual-beam phase calibration based on the horn-on-dish method (Kawaguchi et al.2000) revealed that the instrumental delay in dual-beam system was quite stable, with phase drift less than 3 degree hour$^{-1}$ at 22 GHz.
Therefore, the result presented in this paper would be hardly changed by inclusion of the dual-beam phase calibration.

\section{Results}
\subsection{Dual-Beam Phase Referencing}

Figure 1 shows the residual fringe phase of the intensity peak channel for two maser sources observed on 2002 July 23.
Here the visibilities for W49N and OH43.8-0.1 were integrated for 5 sec, and the constant bias of phase, which is of no interest here, was corrected for so that the residual fringe phase becomes 0 at the time origin (UT 11:30).
In figure 1 the residual fringe phases show strong time-variation due to the atmospheric fluctuations.
During the observed period (3600 sec) the residual phase ranges from -180 to +200 degree.
In particular, from 1300 sec to 1330 sec, the phase varies by 114 degree from -149 to -35 degree.
For fast switching observation, phase jump of $\sim$180 degree is critical because of 360 degree ambiguity of phase, and hence rapid phase variation like this could cause systematic error in fast switching VLBI unless the switching cycle is sufficiently shorter than the time scale of phase variation.

However, the phase variations for W49N and OH43.8-0.1 observed with VERA coincides well with each other.
In fact, the differential phase (difference of two fringe phases in figure 1) is nearly constant at $\sim$0 degree with maximum deviation less than $\pm$30 degrees.
This result strongly demonstrates that the VERA's dual-beam system efficiently calibrates the atmospheric phase fluctuations.
The r.m.s. noise level for the differential phase is 8 degrees.
In the differential phase plot there appear some structures with time scale of 30 to 100 sec (small bumps with peak-to-peak amplitude of 20-30 degrees).
In fact, the power spectrum of the differential phase showed that while the power spectrum is almost flat in most frequency range it has a gentle gradient with power law index of $\sim -1$ in the range of 0.2 Hz to 0.02 Hz (5 to 50 sec in the time domain), confirming the existence of some structures in that time range.
These structures may be due to 1) small-scale structures of atmosphere which was not common to the two sources separated by 0.65 degree and thus was not cancelled out, and 2) structure of maser sources themselves, which also introduce variation in the fringe phase.
We note that the residual differential phase changes moderately when the phase of different channel maser is used, and hence there is indeed some maser structure effect in the differential phase.
In any case, we can smooth out that those structures in the differential phase based on averaging for longer time scale, since the power spectrum of the differential phase is flat beyond time scale of 100 sec.

\subsection{Allan Standard Deviation}

In order to investigate the characteristics of phase variation, we calculated Allan standard deviation for the residual fringe phases and the differential phase.
Allan standard deviation $\sigma_y(\tau)$ can be calculated as follows (e.g., Thompson et al. 2001),
\begin{equation}
\sigma_y^2(\tau) = \frac{\langle\left[\phi(t+2\tau)-2\phi(t+\tau)+\phi(t)\right]^2\rangle}{8\pi^2\nu_0^2\tau^2}.
\end{equation}
Here $\nu_0$ is the observational frequency, $\phi$ is the observed phase, $\tau$ is time interval, and the bracket $\langle \rangle$ denotes the average over the whole samples. 

Figure 2 shows $\sigma_y(\tau)$ for the fringe phases of W49N and OH43.8-0.1 and for the differential phase (both for 0.5 sec integration).
For W49N, $\sigma_y(\tau)$ is nearly constant at $\sim 2\times 10^{-13}$ between $\tau=2$ sec and $30$ sec, indicating that the phase fluctuation is dominated by flicker-frequency noise.
Beyond $\tau=100$ sec, $\sigma_y(\tau)$ for W49N (and also for OH43.8-0.1) becomes almost white-phase noise, decreasing with $\tau^{-1}$.
This feature, a combination of flicker-frequency noise for shorter time scale and white-phase noise for longer time scale, has been well-known as a typical behavior of phase in VLBI (Rogers and Moran 1981; Rogers et al.1984).
In contrast, for OH43.8-0.1 $\sigma_y(\tau)$ for short time scale (less than 10 sec) is proportional to $\tau^{-1}$.
This is due to the S/N limit for OH43.8-0.1, which is fainter than W49N.
In fact, the correlated flux density for OH43.8-0.1 peak channel was 1.8$\times 10^2$ Jy for Mizusawa-Iriki baseline, and an estimate of the Allan standard deviation from the S/N ratio gives 2.3$\times 10^{-12}$ for $\tau=1$ sec, which is in good agreement with the observed value of 2.1$\times 10^{-12}$.

On the other hand, $\sigma_y(\tau)$ for the differential phase is inversely proportional to $\tau$.
This result indicates that the dual-beam phase referencing removed atmospheric fluctuation effectively and that the differential phase is dominated by white phase noise,
We note that in the time range of 5 to 50 sec the differential phase is likely to be dominated by flicker-phase noise due to the phase structure seen in the figure 1, but flicker-phase noise also gives almost the same gradient to the white-phase noise in the Allan standard deviation plot (e.g., Thompson et al.2001), and hence little feature is seen for the differential phase in figure 2.
Since $\sigma_y(\tau)$ for OH43.8-0.1 and $\sigma_y(\tau)$ for the differential phase becomes asymptotically equal toward small time interval ($\tau\le 5$ sec), the differential phase is mainly due to the S/N limit for OH43.8-0.1.

For the data taken on 2002 May 29, we also calculated Allan standard deviation and obtained similar results.
We found that $\sigma_y(\tau)$ for $\tau=100$ sec was $3.7\times 10^{-14}$, instead of $2.9\times 10^{-14}$ on 2002 July 23.
This difference of factor of 1.3 is mainly due to the difference in weather conditions.
Nevertheless, $\sigma_y(\tau)$ for the differential phase taken on 2002 May 29 also shows the linear trend like that in figure 2, proportional to $\tau^{-1}$ .

\begin{figure}
\begin{center}
        \FigureFile(11cm,11cm){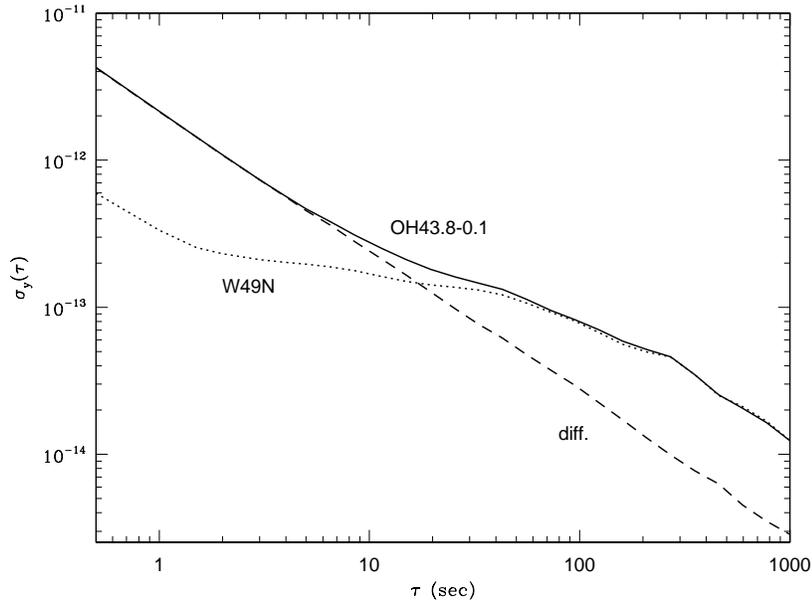}
\end{center}
\caption{Allan standard deviation for fringe phase of W49N (dotted line) and OH43.8-0.1 (thin line), and the differential phase (dashed line).}
\end{figure}

\subsection{Long-Term Integration}

As seen in figure 1, single-beam VLBI strongly suffers from the atmospheric fluctuations.
Due to the rapid phase variation, the complex visibility cannot be integrated beyond so-called 'coherence time', which is around a few minutes at 22 GHz under typical atmospheric condition.
Phase referencing can be a powerful tool to overcome this coherence problem, enabling us to integrate the complex visibility beyond the coherence time.
In order to see how the integration performance is improved based on the dual-beam phase referencing, we calculated the coherence function using the fringe phase and differential phase in figure 1.
The coherence function $C(T)$ can be defined as follows (Rogers and Moran 1981; Thompson et al. 2001),
\begin{equation}
C(T)= \left| \frac{1}{T}\int_0^{T} \exp{\left[\, i \phi(t) \,\right] }\; dt \;\right|,
\end{equation}
where $\phi(t)$ is the observed visibility phase or the differential phase in figure 1, and $T$ is the integration time.

As seen in figure 3, the integration using the fringe phase of W49N is of course ineffective: the amplitude rapidly declines beyond the integration time of 250 sec.
We note that the constancy of the coherence function for W49N beyond 1500 sec is likely to be artificial, mainly due to the {`}high-pass filtering{'} of fringe phase in which slowly-varying long-term drift was corrected for, and real situation would be worse than what appeared in figure 3.
On the other hand, the integration using the differential phase gives an excellent result: the amplitude remains constant for any integration time $t$, and is still 0.98 after 1 hour integration.
This result indicates that based on the phase referencing with VERA's dual-beam system one can integrate the visibility for arbitrarily long period, and hence one can improve the S/N ratio of visibility as $t^{1/2}$.
Thus VERA's dual-beam phase referencing will provide a new and powerful tool to study faint objects that cannot be detected without long-term integration.

\section{Discussion and Conclusion}

Here we discuss whether the phase referencing capability of VERA is sufficient for 10 $\mu$as-level astrometry.
In terms of fringe phase, the target accuracy is $\sim 3$ degree for 22 GHz, which corresponds to 0.1 mm in path length error.
With the maximum baseline of 2300 km (between Mizusawa and Ishigaki-jima stations), one can obtain a rough estimate of VERA's astrometric accuracy as 0.1mm/2300km $\sim$ 10 $\mu$as.
On the other hand, the differential phase residual in figure 1 has the r.m.s. error of $\sim$ 8 degree, which is still larger than the target accuracy of VERA by a factor of 3.
However, we have also seen that the nature of the differential phase is mostly white phase noise that comes mainly from the S/N limit of the fainter source OH43.8-0.1.
Hence integrating the visibility for longer period makes the residual phase error smaller, as phase measurement error $\Delta \phi$ and S/N ratio is related to each other as $\Delta \phi \propto 1/SNR$.
To obtain smaller phase error by a factor of 3, increasing the integration time by an order of magnitude is enough, since the S/N ratio improves with integration time $t$ as $SNR \propto t^{1/2}$.
This indicates that if we integrate the visibility for $\sim$ 50 sec, we can reach down to the differential phase error of 3 degrees.
Thus, the results presented here prove that as far as the calibration of atmospheric fluctuation is concerned, VERA's dual-beam system will allow us to obtain phase accuracy required for 10 $\mu$as-level astrometry.
We note, however, that the pair observed in the present paper had relatively good configurations, having close separation on the sky plane (0.65 degree) and moderate Declination ($\sim +09$d for both W49N and OH43.8-0.1).
Also, the pair sources are sufficiently bright to detect within the coherence time.
For pairs with larger separation and lower Declination, and/or for less bright pairs phase referencing by dual-beam system may not work as efficiently as for the pair in the present paper.
For instance, residual atmospheric phase fluctuation is expected to increase nearly proportionally to the pair separation, with a power law index of 5/6 in case of 'frozen screen' atmosphere with Kolmogorov turbulence (e.g., Carilli and Holdaway 1999).
In order to study how phase referencing capability varies with source configurations as well as weather conditions, we have to observe several pairs in different conditions for several times.
These observations are the next targets of VERA's performance check in near future.
At this stage, however, we may conclude that VERA's dual-beam system can calibrate the atmospheric phase fluctuation with sufficient accuracy for 10 $\mu$as-level astrometry, at least for some bright sources with relatively good configurations.

\begin{figure}
\begin{center}
        \FigureFile(90mm,90mm){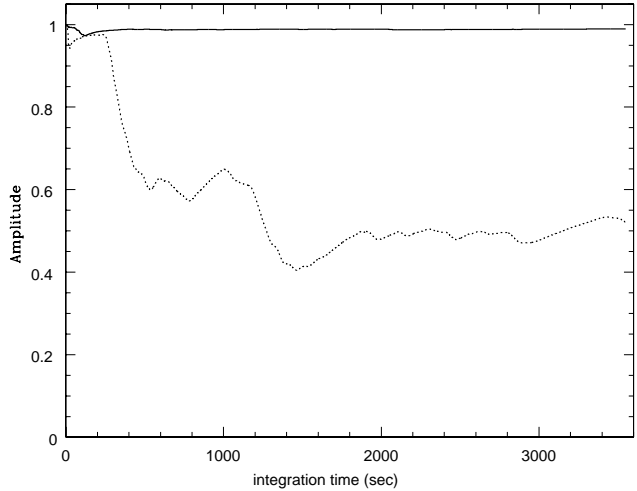}
\end{center}
\vspace{-1.3cm}
  \caption{Coherence function calculated for the fringe phase of W49N (dotted line) and the differential phase (thin line).}
\end{figure}

\vspace{1pc}\par
We are grateful to the referee Dr. James Moran for reviewing and constructive suggestions.
One of the author (MH) acknowledges the financial support from grant-in-aid (No.13740135) from the Ministry of Education, Culture, Sports, Science and Technology.
Part of the data reduction was performed at the Astronomical Data Analysis Center of the National Astronomical Observatory, Japan, which is an inter-university research institute of astronomy operated by Ministry of Education, Science, Culture, and Sports. 


\end{document}